# Valley-based FETs in graphene

M.-K. Lee[1], N.-Y. Lue[1], C.-K. Wen[2], and G. Y. Wu[1,2,*]

[1]Department of Physics, National Tsing-Hua University, Hsin-Chu 30013, Taiwan, ROC;
[2]Department of Electrical Engineering, National Tsing-Hua University, Hsin-Chu 30013, Taiwan, ROC; [*]E-mail: yswu@ee.nthu.edu.tw.

**Abstract**

An analogue of the Datta-Das spin FET is investigated, which is all-graphene and based on the valley degree of freedom of electrons / holes. The "valley FET" envisioned consists of a quantum wire of gapped graphene (channel) sandwiched between two armchair graphene nanoribbons (source and drain), with the following correspondence to the spin FET: valley (K and K') ↔ spin (up and down), armchair graphene nanoribbons ↔ ferromagnetic leads, graphene quantum wire ↔ semiconductor quantum wire, valley-orbit interaction ↔ Rashba spin-orbit interaction. The device works as follows. The source (drain) injects (detects) carriers in a specific valley polarization. A side gate electric field is applied to the channel and modulates the valley polarization of carriers due to the valley-orbit interaction, thus controlling the amount of current collected at the drain. The valley FET is characterized by: i) smooth interfaces between leads and the channel, ii) strong valley-orbit interaction for electrical control of drain current, and iii) vanishing interband valley-flip scattering. By its analogy to the spin FET, the valley FET provides a potential framework to develop low-power FETs for graphene-based nanoelectronics.

# I. INTRODUCTION

The pioneering work of Datta and Das in semiconductor spin FETs[1] has opened a door to the utilization of spin degree of freedom (d.o.f.) for the control of electrical transport in semiconductors, and has inspired many important research ideas, in addition to those based on GMR or TMR, for spintronic applications[2]. The realization of such FETs will not only permit the spin to be used as a logic variable, but also make it possible to further down-scale the semiconductor transistor size due to the low power consumption required for spin FETs.

The prototype FET considered by Datta and Das is a simple structure, consisting of a ballistic, quasi-one-dimensional (Q1D) channel made of a semiconductor with large Rashba spin-orbit interaction (SOI), and ferromagnetic (FM) leads as the source and drain. FM leads inject and detect spin-polarized electrons with the spin orientation determined by the magnetization in the leads. For current control, a (top) gate electric field is applied to the channel and generates an effective magnetic field due to the Rashba SOI,

$$H_{so} = \alpha\, \boldsymbol{\sigma} \cdot (\boldsymbol{k} \times \bar{\boldsymbol{e}})$$

($\alpha$ = Rashba constant, $\boldsymbol{\sigma}$ = Pauli matrix, $\boldsymbol{k}$ = electron wave vector, and $\bar{\boldsymbol{e}}$ = directional vector of the electric field). $H_{so}$ induces a spin precession, thus controlling the spin orientation of channel electrons and hence the current collected at the drain. Detection of the Rashba effect has been demonstrated recently in the case of a two-dimensional electron gas.[3]

The Datta-Das spin FET permits the modulation of conductance via electrical manipulation of spins. However, its realization has met major challenges. Among the important issues studied for spin FETs are i) low injection / detection efficiency for diffusion-based current injection, due to the conductivity mismatch between the FM leads and the channel semiconductor;[4] and ii) random conductance oscillations resulting from the SOI-induced ballistic interband spin-flip scattering.[5] Various solutions have been attempted. For example, the insertion of a tunnel junction between the electrode and the semiconductor is proposed[6] to resolve i), and using stray electric fields for obtaining a reasonable spin control to resolve ii).[7]

On the other hand, the recent rise of the wonder material, graphene,[8-10] provides a novel road to the future FETs. Being a two-dimensional sheet of carbon atoms with excellent carrier mobility, graphene offers the thinnest possible channel, and the possibility to scale to shorter channels and higher speeds for MOSFETs.[11] More importantly, it also provides an unprecedented flexibility to the design of nanoelectronic devices, due to the expanded family of electron d.o.f.s introduced. Because graphene has two degenerate and inequivalent energy valleys (K and K'), carriers in it are endowed with the extra character − valley, besides spin and charge, for information processing. This leads to the emergence of a new category of electronics known as "valleytronics" which manipulates the valley d.o.f. for control of electronic properties.[12] Implementation of prototype devices such as valley filters[12] or valley-based qubits for quantum computing[13] / communications[14] have recently been demonstrated theoretically.

In this work, we discuss a feasible valleytronic implementation of the Datta-Das idea, for graphene-based electronics. The proposed "valley FET" (VFET) consists of a Q1D channel of gapped graphene sandwiched between two armchair graphene nanoribbons (AGNR) (source and drain). Being all-graphene, the VFET is free from the problem of interface scattering and energy band mismatch (i.e., Issue i) above). Moreover, it employs the physical mechanism, the so-called valley-orbit interaction (VOI), for the electrical control of drain current. The mechanism exists uniquely in gapped graphene and is similar to the Rashba SOI, with a significant difference, though. As derived previously, it is given by[13-15]

$$H_{vo} = \tau \frac{\hbar}{4m^*\Delta}(\nabla V \times \vec{p}) \cdot \hat{z}$$

which is valley-conserving ($\tau = +/-$ being the valley index for K / K', $2\Delta$ = energy gap, $m^*$ = electron effective mass = $\Delta / v_F^2$, $v_F$ = Fermi velocity, $V$ = potential energy, $\vec{p}$ = momentum operator, $\hat{z}$ = unit vector normal to the graphene plane). Therefore, the VOI does not induce the flip type scattering which causes complications such as Issue ii). Without Issues i) and ii), the VFET provides a promising implementation of the Datta-Das idea.

The presentation is organized as follows. In Sec. II, the structure of VFETs is proposed, and the theory underlying VFETs is discussed. In Sec. III, summary of this work is given and directions for future studies are suggested.

## II.  Valley FETs

## Structure

The structure of a valley FET is shown in Fig. 1. The source and drain are made of AGNRs. The channel section is aligned with the armchair direction, with the zigzag edges of the region being passivated for stabilization.[16] The channel is subject to an electric potential (due to, for example, a back gate bias) for lateral electron confinement as well as a (side gate) electric field for conductivity modulation. An h-BN[17,18] or SiC[19] substrate may be used to grow the structure and open an energy gap ($2\Delta$) in the structure. In the case of h-BN grown graphene, the slight lattice constant mismatch (~ 1.8%) between graphene and BN results in the formation of Moire pattern[18] and a corresponding periodic variation in space in the gap parameter $\Delta$. In order to avoid this complication, we assume that a biaxial strain is applied upon either graphene or BN to ensure lattice match, giving a uniform $\Delta$ throughout the structure. With the choice of Cartesian coordinates shown in the figure, the gap occurs at the points **K** = $(0, 4\pi/3a_0)$ and **K'** = $(0, -4\pi/3a_0)$ of the Brillouin zone ($a_0$ = graphene lattice constant).

The table below shows the close correspondence between spin and valley FETs, in their structures as well as principles of operation:

| **FET** | **d.o.f.** | **lead** | **Q1D channel** | **electrical control** | **physical mechanism** |
|---|---|---|---|---|---|
| valley FET | valley K,K' | AGNR | graphene | side gate | valley-orbit interaction (VOI) |
| spin FET | spin ↑,↓ | FM | semiconductor | top gate | Rashba SOI |

In a VFET, the source and drain AGNRs polarize electrons in a specific valley state for injection / detection, and the (side) gate field is applied to the channel and induces "valley precession" of channel electrons for the control of drain current, due to the VOI. We discuss the lead state, channel state, and valley precession in the following.

## Lead states

Firstly, we consider the valley polarization of electrons in the source and drain. As shown previously,[20] K and K' valleys are mixed in an AGNR cut out of gapless graphene ($\Delta = 0$). This is also true in the case with $\Delta \neq 0$. Let W = the ribbon width and $\psi_{D,\tau} = (\psi_{A,\tau}(\mathbf{r}), \psi_{B,\tau}(\mathbf{r}))^T$ be the $\tau$ valley, two-component Dirac wave function describing the electron probability amplitudes on A and B sites of the graphene crystal, respectively. $\psi_{D,\tau}$ satisfies the following Dirac type equation[10]

$$\begin{pmatrix} \Delta & v_F(i\hat{p}_x + \tau \hat{p}_y) \\ v_F(-i\hat{p}_x + \tau \hat{p}_y) & -\Delta \end{pmatrix} \begin{pmatrix} \psi_{A,\tau} \\ \psi_{B,\tau} \end{pmatrix} = (E + \Delta) \begin{pmatrix} \psi_{A,\tau} \\ \psi_{B,\tau} \end{pmatrix}. \quad (1)$$

The electron energy E here is measured with respect to the conduction band edge. We take E > 0 (i.e., the electron case) throughout this work. The hole case can be similarly treated due to the electron-hole symmetry in Eqn. (1).

For an electron traveling down the nanoribbon with wave vector $k$, the real space wave function is a linear combination,

$$\begin{pmatrix} \Psi_A \\ \Psi_B \end{pmatrix} = e^{i\vec{K}\cdot\vec{r}} \psi_{D,+} + e^{i\vec{K}'\cdot\vec{r}} \psi_{D,-},$$

subject to the boundary conditions $\Psi_A(r) = \Psi_B(r) = 0$ at the edges of the ribbon (located at y = ±W/2). The solution is given by

$$\begin{pmatrix} \Psi_A \\ \Psi_B \end{pmatrix} \propto \left[ \begin{pmatrix} e^{ikx} e^{i\vec{K}\cdot\vec{r}} & S_{K'/K} e^{ikx} e^{i\vec{K}'\cdot\vec{r}} \end{pmatrix} \begin{pmatrix} e^{ik_y y} \\ e^{-ik_y y} \end{pmatrix} \right] \begin{pmatrix} 1 \\ \dfrac{\hbar v_F(k_y - ik)}{2\Delta + E} \end{pmatrix}, \quad (2)$$

which is laterally quantized, with $S_{K'/K} = (-1)^{n+1}$ and the quantized values $E = E_n$ and $k_y = k_n$, where $(E_n+\Delta)^2 = \Delta^2 + \hbar^2(k^2 + k_n^2)/2m^*$ and $k_n = n\pi/W - 4\pi/3a_0$, respectively. Here, $S_{K'/K}$ is the amplitude of K' component relative to that of K component. Therefore, the source / drain state in Eqn. (2) is valley-mixed in a 50-50 ratio. This specific "valley polarized state" is utilized for valley injection / detection, in parallel to its spin counterpart in the Datta-Das case, where the spin injected / detected is specifically aligned in the channel direction (e.g., "→") to have even amounts of "↑" and "↓" components.

**Channel states**

Next, we consider the state of electrons in the Q1D channel of gapped graphene. The Dirac equation in the region is given by Eqn. (1), with the following potential energy

$$V(y) = \frac{m^* w_0^2 y^2}{2} - Dy^4 + e\varepsilon_y y \quad (D > 0)$$

added to the diagonal elements of the equation. V(y) consists of three terms and describes both the lateral confinement and the (side) gate potential for the Q1D channel − the 1st (parabolic) and the 2nd (quartic) terms combined to represent the confinement potential which is parabolic near y ~ 0 but flattens out for $|y| \sim O[(m^*/D)^{1/2} w_0]$, and the 3rd (linear) term being the gate-induced electric potential energy ($\varepsilon_y$ = side gate electric field). Note that the inclusion of negative $Dy^4$ in V(y) is realistic and describes the case of a finite confinement potential, which eventually flattens out at large y.

We focus on the low-energy regime where the Fermi energy (or E) is located near the conduction band edge. In the absence of V(y), solving the Dirac equation yields $(E+\Delta)^2 = \Delta^2 + v_F^2 \hbar^2 (k^2 + k_y^2) / 2\Delta$, where $(k, k_y)$ = wave vector of the particle. This is the standard dispersion of a free massive Dirac particle in two dimension, with $\Delta$ = "rest mass energy" and $v_F$ = "light speed".[10] In the presence of V(y), the Dirac equation is difficult for analytical treatment. However, for $E \ll \Delta$, the quantum mechanics of electrons belongs to the "non-relativistic regime" and is well described in the framework of "Schrodinger description", an excellent approximation to the Dirac description as shown previously.[13,14] We follow this framework below, and employ the simple, one-component "Schrodinger wave function" for the discussion of channel electrons. The wave function satisfies

$$H\phi_{n,\tau} \approx E_{n,\tau} \phi_{n,\tau},$$
$$H = H^{(0)} + H^{(1)}, \quad (3)$$
$$H^{(0)} = \frac{\vec{p}^2}{2m^*} + V, \quad H^{(1)} = -\frac{\vec{p}^4}{8m^{*2}\Delta} - \frac{\vec{p}^2 V}{8m^*\Delta} + H_{vo},$$

H is the Schrodinger Hamiltonian, with $H^{(0)}$ being the "non-relativistic part" and $H^{(1)}$ the "1st-order relativistic correction (R.C.)". $E_{n,\tau}$ is the energy level with lateral quantum index n, for $\tau$-valley electrons. $\phi_{n,\tau}$ is the corresponding Schrodinger wave function and derives from the component $\psi_{A,\tau}$ of Dirac wave function by the linear transformation, $\phi_{n,\tau} = (1 + p^2 / 8m^*\Delta) \psi_{A,\tau}$. Specifically, $\phi_{n,\tau}$ is interpreted as a probability amplitude, with $|\phi_{n,\tau}|^2 \approx |\psi_{A,\tau}|^2 + |\psi_{B,\tau}|^2$ (i.e., the probability distribution of an electron over unit cells). The close analogy between the present description and the standard Schrodinger quantum mechanics (with relativistic effects included) is obvious in Eqn. (3). Apart from the usual R.C. "$-p^4/8m^{*2}\Delta$" to the kinetic energy and the Darwin term "$-p^2V/8m^*\Delta$", $H_{vo}$ appears in $H^{(1)}$ as well, in the same way as $H_{so}$ does as a part of R.C. to the standard Schrodinger theory. $H_{vo}$ couples valley and orbit, making it possible to manipulate the valley d.o.f. by electric means.

Eqn. (3) can be solved analytically within the perturbation theory, in the limit of V(y) being dominantly parabolic, where the simple harmonic oscillator (SHO) states may be used for the calculation. In the following, we state the conditions for the analysis to be valid. First, two energy scales are involved in the problem, namely, $\hbar w_0$ (SHO energy) and $\Delta$, for which we require

a) $\hbar w_0 \ll \Delta$ (non-relativistic regime).

We further impose the following conditions on various energies:

b) $\|p_x^2/2m^*\| < \hbar w_0$;

c) $\|Dy^4\| \ll \hbar w_0$ and $\|e\varepsilon_y y\| \ll \hbar w_0$ (weak linear and quartic potentials).

To facilitate the perturbation-theoretical calculation, the Hamiltonian is organized into the following form,

$$H = H_0 + H',$$
$$H_0 = \frac{\vec{p}^2}{2m^*} + \frac{1}{2}m^* w_0^2 y^2,$$
$$H' = -Dy^4 + e\varepsilon_y y - \frac{\vec{p}^4}{8m^{*2}\Delta} - \frac{\vec{p}^2 V}{8m^*\Delta} + H_{vo}.$$

$H_0$ is the Hamiltonian of unperturbed system and $H'$ is the total perturbation. Given Conditions a)-c) above, it can be shown that $\|H'\| \ll \|H_0\|$. The eigenstates and eigenvalues of $H_0$ are given by

$$\phi_{n,\tau}^{(0)}(x,y;k_\tau) = e^{ik_\tau x} <y|n>,$$
$$E_{n,\tau}^{(0)}(k_\tau) = (n+\frac{1}{2})\hbar w_0 + \frac{\hbar^2 k_\tau^2}{2m^*},$$

where $\phi_{n,\tau}^{(0)}$ = wave function, $E_{n,\tau}^{(0)}$ = energy, $k_\tau$ = the electron wave vector (for valley index $\tau$) along the channel, and $|n\rangle$ = the SHO eigenstate with quantum index n.

Now, we carry out the perturbation-theoretical calculation. To keep the discussion simple, we focus on the one-channel case, where only the lowest subband (n = 0) is occupied and utilized for transport.[21] Let $\delta E$ = valley-dependent energy correction due to H'.[22] The leading-order contribution to $\delta E$ is given by the following 2$^{nd}$-order perturbation-theoretical expression,

$$\delta E \approx -\sum_m{}' \frac{<0|e\varepsilon_y y|m><m|H_{vo}(due\ to\ the\ quartic\ potential)|0>}{m\hbar w_0} + c.c.,$$

$$H_{vo}(due\ to\ the\ quartic\ potential) = \tau \frac{\hbar^2}{m^*\Delta} k_\tau D y^3.$$

The above expression for $\delta E$ is evaluated analytically, giving

$$\delta E = -\alpha_{vo} \tau\, k_\tau, \qquad (4)$$
$$\alpha_{vo} = \frac{3e\hbar^3}{2m^{*3} w_0^3 \Delta} D\varepsilon_y.$$

For an electron with wave vector $k_\tau$, we summarize the perturbation-theoretical results for $E_{0,\tau}$ and (unnormalized) $\phi_{0,\tau}$ in the following

$$E_{0,\tau}(k_\tau) \approx \frac{1}{2}\hbar w_0 + \frac{\hbar^2 k_\tau^2}{2m^*} - \alpha_{vo}\tau\, k_\tau, \qquad (5)$$
$$\phi_{0,\tau} \approx e^{ik_\tau x}\exp(-\beta y^2).$$

The Gaussian function with the parameter $\beta = m^* w_0 / 2\hbar$ is the (unnormalized) SHO ground state. Notice the presence of a "Rashba term", i.e., "$\alpha_{vo}\tau\,k_\tau$" in the subband dispersion $E_{0,\tau}(k_\tau)$, with $\alpha_{vo}$ being the corresponding "Rashba constant". In the case where D = 0, the Rashba term vanishes. This can be understood since the linear (electric) potential term in V(y) can be made to disappear by making an y-coordinate shift: $y \to y' = y + y_\varepsilon$, where $y_\varepsilon = e\varepsilon_y / m^* w_0^2$, giving $V(y) \approx \frac{1}{2} m^* w_0^2 y'^2$ (correct to $O(y_\varepsilon)$). Eqns. (4) and (5) constitute the main result of this work.

$E_{0,\tau}(k_\tau)$ is plotted in Fig. 2. We see that the subbands for K and K' are horizontally split due to the Rashba term. For a given energy E, the splitting is given by

$$k_+ - k_- = \frac{2m^*\alpha_{vo}}{\hbar^2} \tag{6}$$

independent of the energy.

The magnitude of $\alpha_{vo}$ reflects the strength of the VOI. $\alpha_{vo}$ is estimated as follows. We take $v_F = 10^6$ m/s,[10] $\Delta = 0.026$ eV (for graphene grown on h-BN)[17] and $\varepsilon_y = 2.5\mu$V/Å. For the parabolic potential parameter $w_0$, we take $\hbar w_0 / \Delta = 0.4$. With this choice, the effective Q1D channel width (defined as $2\beta^{-1/2}$) = 1100Å. We take the quartic potential parameter $D = mw_0^2\beta / 2$. At this value, the confinement potential flattens out near $x = \pm\beta^{-1/2}$, i.e., the edges of the Q1D channel. With these parametric values, Eqn. (4) yields[23]

$$\alpha_{vo} \approx 6.4 \times 10^{-12} eV \cdot m,$$

which is comparable to the Rashba constant in semiconductors with large SOI, e.g., InAs.[24]

**Valley precession**

The channel state of an injected electron is a linear combination of K and K' components, with the following real space wave function

$$\Phi_0 \approx \begin{pmatrix} e^{ik_+x}e^{i\vec{K}\cdot\vec{r}} & C_{K'/K}e^{ik_-x}e^{i\vec{K}'\cdot\vec{r}} \end{pmatrix}\begin{pmatrix} \exp(-\beta y^2) \\ \exp(-\beta y^2) \end{pmatrix}. \tag{7}$$

The parameter $C_{K'/K}$ here is the amplitude of K' component relative to that of K component, and is chosen to match the valley polarization of channel state to that of source state (as specified in Eqn. (2)) at the channel / source interface (located at $x = 0$). This gives $C_{K'/K} = S_{K'/K}$. Therefore, K and K' valleys are again mixed in the 50-50 ratio, as in the case of lead states. But there is an important difference. In contrast to the lead case, where the phases of K and K' components evolve with the same wave vector (i.e., k in Eqn. (2)), now they evolve separately with different wave vectors ($k_+$ and $k_-$), due to the Rashba effect discussed earlier. This leads to the valley precession of channel electrons in the valley space. Specifically, after the electron travels for a distance L (L = channel length), the phase difference between the two valley components is given by

$$\delta\varphi = \frac{2m^*\alpha_{vo}}{\hbar^2}L. \tag{8}$$

δφ determines the orientation of valley polarization before the electron enters the drain, relative to that of the drain state. For δφ = 2mπ, the two polarizations are aligned, leading to a conductance maximum. On the other hand, for δφ = (2m+1)π, they are orthogonal to each other, leading to a conductance minimum. Since δφ scales linearly with the gate electric field $\varepsilon_x$ (via the dependence of $\alpha_{vo}$ on $\varepsilon_x$), it gives a VFET the on-off switch capability through gate control. Note also that since δφ as given by Eqn. (8) is independent of electron energy (within the accuracy of our analysis), a VFET shares the nice characteristics of a spin FET, in that the on-off swing is insensitive to the spread in electron energy distribution, an important requirement for a good on-off ratio.[1]

**Effects of short-range impurity potentials**

Last, we briefly discuss impurity effects on the performance of a VFET. In general, in a nanoscale FET, impurities in the channel block the carrier transport and cause detrimental effects on the device performance. In the case of a VFET, the impurities with short-range potentials are of particular concern, because they provide the large wave vector difference needed in the valley flip scattering K ↔ K′ and thus may damage the valley precession. In the following, we consider such impurities and compare the valley flip time ($T_{flip}$) to the transit time ($T_{transit}$), for an electron moving through the channel.

Substituting in Eqn. (8) δφ = π and the values of $v_F$, Δ (for $m^*$) and $\alpha_{vo}$ given earlier, we estimate L ~ O(μm) as the channel length required (for the on-off switch function, in the VFET where the gate field $\varepsilon_y$ is specifically taken to be 2.5μV/Å). We further take the electron kinetic energy along the channel to be O($\hbar w_0$). This gives the ballistic transit time

$$T_{transit} = O\left(\frac{L}{\sqrt{2\hbar w_0 / m^*}}\right) = O(10^{-12} \text{ sec}).$$

For the estimate of $T_{flip}$, we take the total impurity potential to be

$$V_{im}(x,y) = a_{im}^2 V_{0,im} \sum_{n=1}^{N_{im}} \delta(\vec{r} - \vec{R}_n),$$

where $a_{im} = O(\text{Å})$ being the range of each impurity potential, $V_{0,im} = O(\text{eV})$ being the potential strength, "$R_n$" = impurity position vector, and $N_{im}$ = total number of impurities in the channel. Application of Fermi's Golden Rule yields the following valley flip rate due to the impurities

$$\frac{1}{T_{flip}} = O\left(N_{im} \frac{2\pi}{\hbar} V_{0,im}^2 \frac{a_{im}^4}{L^2 W_c^2} D_{1d}\right) = O(N_{im} \cdot 10^6 \text{ sec}^{-1}),$$

where $D_{1d}$ is the electron density of states in one dimension, given by $D_{1d} = O\left(\frac{L}{\pi\hbar}\sqrt{\frac{m^*}{\hbar w_0}}\right)$, and $W_c = 2\beta^{-1/2} = 1100$Å being the channel width.

Overall, we obtain the ratio

$$T_{transit} / T_{flip} = O(N_{im} \cdot 10^{-6}),$$

which is generally very small. The ratio becomes $O(1)$ only if we extrapolate the result to the extremely dirty limit where $N_{im} = O(10^6)$. (In the specific VFET considered here where the channel size is characterized by $L = O(\mu m)$ and $W_c = 1100$Å, it means all or a substantial fraction of channel atoms are impurities.) Therefore, as far as the valley precession is concerned, the above estimate indicates that the precession is completed long before the valley flips. As such, the scattering by short-range impurity potentials does not pose any serious problem.

### III. Summary and Future Work

In summary, we have investigated the feasibility of a valleytronic implementation of the Datta-Das idea. The device envisioned is all graphene, with AGNRs as electrodes and a graphene quantum wire as the channel. Moreover, the VFET is free of the issues concerning injection / detection efficiency or interband valley flip scattering, and by its analogy to the spin FET, provides a potential framework to develop low-power FETs for graphene-based nanoelectronics.

Throughout the work, we have ignored the spin d.o.f. of electrons and the effect of Rashba SOI on electron transport, which is known to be extremely weak in graphene compared to that in a typical semiconductor.[25] Instead, we have proposed to employ the VOI for electrical control of graphene FETs, which operates on the valley d.o.f. and has the characteristic of being valley-conserving. With the large magnitude of Rashba constant - $\alpha_{vo}$ - shown in this work, the VOI is established as the dominant mechanism for electrical control of valley FETs.

Finally, the present work has analyzed the VFET within the simple formalism of perturbation theory. As such, only a limited scope of VFET physics has been covered. Various important issues, such as varied implementations, defect scattering, multi-channel transport, the field effect at large gate bias or in the relativistic regime, or the wave interference due to the presence of lead / channel interfaces, are worth further investigations, in order to provide extensive insights into VFETs.

**Acknowledgement** – We thank the support of ROC National Science Council through the contract No. NSC100-2112-M-007-009.

21. We assume that the leads and channel are biased in such a way as to approximately align the lowest subbands in these different regions.
22. Valley-independent energy correction is ignored here, because it does not contribute to the wave vector difference (shown in Eqn. (6)) and hence has no effects on the valley precession.
23. For the order-of-magnitude estimate, we expect the expression in Eqn. (4) for $\alpha_{vo}$ to be valid for the values of $w_0$ and D considered here.
24. A. Wirthmann, Y.S. Gui, C. Zehnder, D. Heitmann, C.-M. Hu, and S. Kettemann, Physica E **34**, 493 (2006).
25. H. Min, J. E. Hill, N. A. Sinitsyn, B. R. Sahu, L. Kleinman, and A. H. MacDonald, Phys. Rev. B **74**, 165310 (2006).

**Figure Captions**

**Fig. 1. (a)** The VFET shown as a three-terminal device. The source and drain are AGNRs, which inject and detect electrons in a specific polarization. The Q1D channel is a quantum wire of gapped graphene, subject to the (side) gate bias. When an electron moves down the channel, the valley polarization vector of the electron precesses due to the VOI. **(b)** The corresponding graphene crystal structure of the device, with the channel region being subject to a lateral confinement potential in order to form a Q1D channel, and the zigzag edges of this section being passivated for stabilization.

**Fig. 2.** $E_{0,\tau}(k_\tau)$ is plotted. For a given energy E, the subbands for K and K' are horizontally split due to the Rashba term in Eqn. (5).

(a)

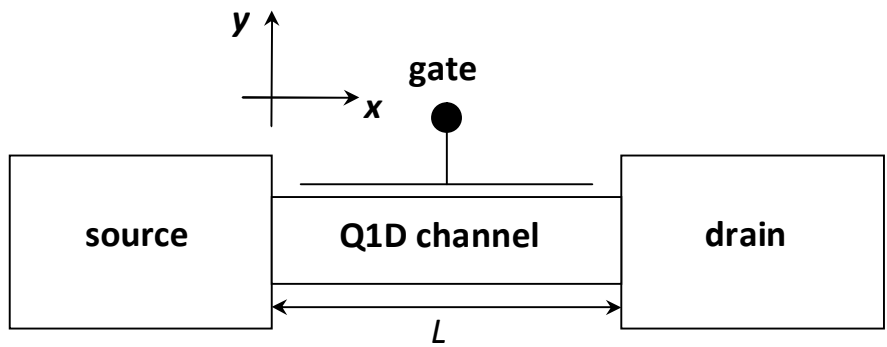

(b)

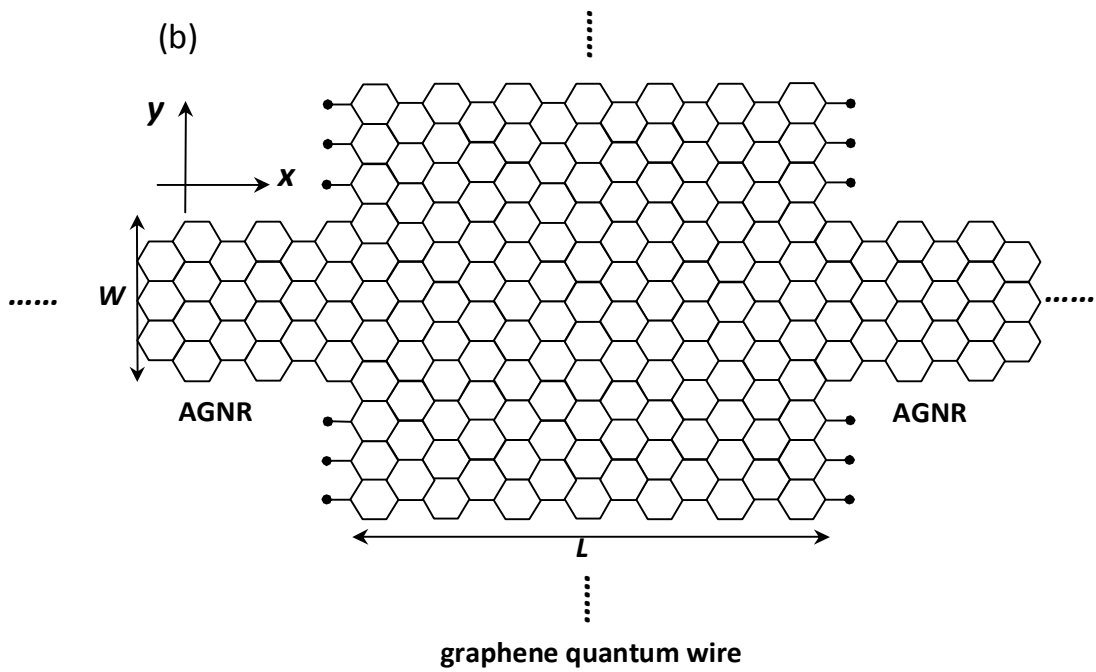

**Figure 1**

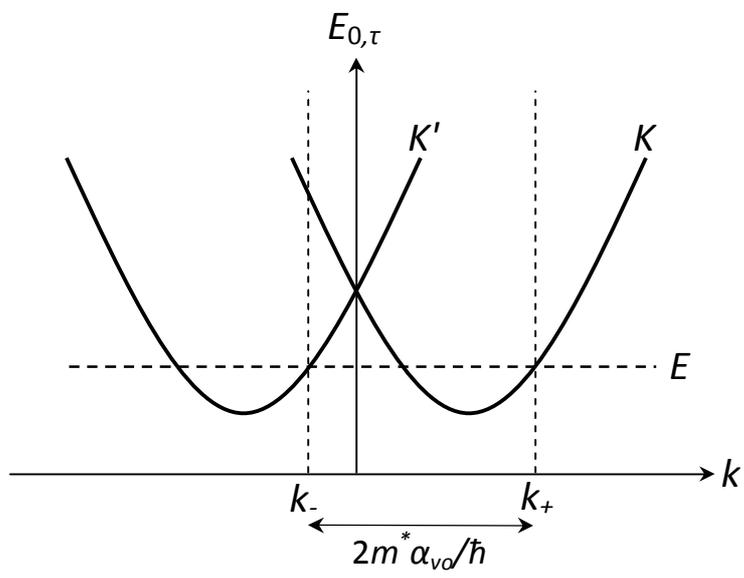

**Figure 2**

# Erratum: Valley-based FETs in graphene [Phys. Rev. B 86, 165411 (2012)]

N.-Y. Lue[1], Y.-C. Chen[2], and G. Y. Wu[1,2,*]

[1] Department of Electrical Engineering, National Tsing-Hua University, Hsin-Chu, Taiwan 30013, ROC

[2] Department of Physics, National Tsing-Hua University, Hsin-Chu, Taiwan 30013, ROC

[*] Corresponding author; Email: yswu@ee.nthu.edu.tw

A mistake occurs in Eqn. (4) for the estimate of $\alpha_{vo}$ in the FET channel, and should be corrected as follows. In the calculation of the valley splitting energy, $\delta E$, we have missed contributions from the 3$^{rd}$-order perturbation theory, e.g.,

$$\sum_{n,m}{}' \frac{<0|e\varepsilon_y y|n><n|(\text{quartic potential})|m><m|H_{vo}(\text{due to the quadratic potential})|0>}{(-n\hbar w_0)(n-m)\hbar w_0} \text{ etc.}$$

In comparison to the original (2$^{nd}$-order perturbation-theoretical) expression given for $\delta E$, these additional contributions have the same dependence in $\varepsilon_y$, D, $k_x$, and $\Delta$, as the 2$^{nd}$-order expression. More importantly, it is found that the overall 3$^{rd}$-order contributions cancel exactly with the 2$^{nd}$-order expression, yielding a vanishing $\delta E$, to the 1$^{st}$-order relativistic correction.

However, as shown below, $\delta E$ becomes finite, if we introduce a variation in the gap in the y-direction, e.g., $\Delta(y) = \Delta + \Delta'(y)$, where $\Delta'(y)$ describes the variation. $\Delta'(y)$ may be piecewise constant, e.g., $\Delta'(y) = 0$ in the channel, and $\Delta'(y) = \Delta_0'$ outside the channel. (See the discussion at the end, for a feasible realization of a gap varying structure.) In the following calculation, we model $\Delta'(y)$ with a quadratic function, e.g., $\Delta'(y) = \delta\, y^2$, where $\delta << m^* w_0^2$ (i.e., weak gap variation).

First, on account of gap variation in this case, the Schrodinger equation in (3) is modified as follows:

$$\left(H^{(0)} + H_{vo}' + H_0^{(1)}\right)\phi_{n\tau} = E_{n\tau}\phi_{n\tau},$$

$$H^{(0)} = \frac{\vec{p}^2}{2m^*} + [V(y) + \Delta'(y)], \qquad (3')$$

$$H_{vo}' = H_{vo} - \tau \frac{\hbar}{4m^*\Delta}[\nabla\Delta'(y) \times \vec{p}] \cdot \hat{z}.$$

$H_0^{(1)}$ here denotes the valley-independent relativistic correction, and is irrelevant to the present discussion. Notice in the above equation the modification of valley-orbit interaction, e.g., $H_{vo} \to H_{vo}'$, in the presence of $\Delta'(y)$. Moreover, in order to facilitate the calculation, we shift the y-coordinate, e.g., $y \to y + y_\varepsilon'$ ($y_\varepsilon' \equiv e\varepsilon_y / (m^* w_0^2 + 2\delta)$), and write

$$V(y) + \Delta'(y) \approx \frac{m^* w_0^2 y^2}{2} - Dy^4 + 4Dy_\varepsilon' y^3 + \delta\, y^2,$$

where we have ignored the terms which are $O(y_\varepsilon'^2)$.

$\delta E$ is given by the following first-order perturbation-theoretical expression due to $H_{vo}'$,

$$\delta E = -\frac{\tau\hbar^2}{4m^*\Delta} k_x <\phi_{0\tau}^{(0)} | \{\partial_y[V(y) - \Delta'(y)]\} | \phi_{0\tau}^{(0)}>,$$

$$E_{n\tau}^{(0)}\phi_{n\tau}^{(0)} = H^{(0)}\phi_{n\tau}^{(0)}.$$

To the 1st order in $\delta$, this yields the following results

$$\delta E \approx -\tau \frac{6e\hbar^3}{m^{*4}\Delta_0 w_0^5} D\delta\,\varepsilon_y k_x, \qquad (4')$$

$$\alpha_{vo} = \frac{6e\hbar^3}{m^{*4}\Delta_0 w_0^5} D\delta\,\varepsilon_y.$$

which show the same dependence in D, $\varepsilon_y$, $\Delta$, and $k_x$ as the previous $\delta E$ and $\alpha_{vo}$ in Eqn. (4).

We estimate $\alpha_{vo}$ with the same parameters (e.g., D, $\hbar w_0/\Delta$, $\varepsilon_y$) used earlier, and take $\delta = 0.2\beta\Delta$. This gives $\alpha_{vo} \sim 5.4 \times 10^{-12}$ eV-m, which is of the same order of magnitude as the previous estimate.

Last, we note that the condition of gap variation in the present discussion can be realized, in principle, by at least two methods, as follows. **I)** In the case of monolayer graphene grown on h-BN, one can make a trough in the BN substrate, and place the graphene layer upon the substrate. While the (strip) region of graphene on top of the trough is free-standing and gapless, the graphene-substrate interaction generates a gap in graphene next to the trough. This structure provides not only a gap variation but also a gap-caused quantum confinement for electrons in the strip, resulting in a quantum wire. **II)** In the case of bilayer graphene, a gap can be generated by applying a DC bias across the two layers [1], and a variation in the DC bias produces a varied

band gap in graphene.[2]